\documentclass[aps,prl,twocolumn,superscriptaddress]{revtex4}
\usepackage{graphicx}%        Include       figure       files
\usepackage{amsmath,amssymb}
\begin{document}

\title{Beam Instabilities in the Scale Free Regime}

\author{V. Folli} 
\affiliation{Institute for Complex Systems - CNR, Dep. Physics University Sapienza,  Piazzale Aldo Moro 2, 00185 Rome, Italy}
\author{E. DelRe} 
\affiliation{Department of Physics, University Sapienza, P.le Aldo Moro 5, 00185, Rome (IT)}
\author{C.Conti} 
\affiliation{Department of Physics, University Sapienza, P.le Aldo Moro 5, 00185, Rome (IT)}
\affiliation{Institute for Complex Systems - CNR, Dep. Physics University Sapienza,  Piazzale Aldo Moro 2, 00185 Rome, Italy}
\date{\today}
%\pacs{...}

\begin{abstract}
The instabilities arising in a one-dimensional beam sustained by the diffusive photorefractive nonlinearity in out-of-equilibrium ferroelectrics are theoretically and numerically investigated. In the ``scale-free model'', in striking contrast with the well-known spatial modulational instability, 
two different beam instabilities dominate: a defocusing and a fragmenting process.
Both are independent of the beam power and are not associated to any specific periodic pattern.
\end{abstract}
\maketitle

%%%%%%%%%%%%%%%%%
\noindent {\it Introduction ---}
Instability drives optical nonlinear waves into new and unexpected regimes.
As for emergent behaviors in complex systems, so instability can lead a simple and featureless input wave
to a whole spectrum of effects that range from apparently random fragmentation to highly
regular pattern formation.
Systems supporting solitons manifest a very general form of instability,
modulational instability (MI), which fragments beams into patterns of periodic spots.\cite{Campillo73, Law1993, Mamaev96, Fuerst97, Shiek01, Peccianti03,DelRe09book}
The entire process is rooted in the presence of one dominant spatial scale 
that forms in the nonlinear interaction, and this scale drives sufficiently wide 
waves into periodic patterns with a precise spatial period.

Recently, the observation of a novel class of optical spatial solitons \cite{KivsharBook,TrilloBook}
has been reported in disordered out-of-equilibrium ferroelectrics \cite{DelRe11, Conti11}.
In this kind of media, the diffusion-driven photo-refractive nonlinearity \cite{Crosignani99} can be largely enhanced to support ``scale-free'' non-diffractive 
beams, which can have arbitrary amplitude and waist.
The mechanism is a direct by-product of the photorefractive band-transport model in disordered systems, and applies to all photorefractive crystals, such as SBN and BaTiO$_3$, that are previously depoled into ferroelectric clusters (see, for example, \cite{Chauvet2006}), alongside organic photorefractive glass \cite{Asaro2005} and photorefractive polymers \cite{Sheu2001}.  More generally, scale-free optical effects stem from the interplay between diffraction and diffusion that mingle to cancel out their respective and intrinsic spatial scales, a condition that can occur in any wave supporting systems where diffusion and diffraction both play a relevant role. One known example is in slow-light experiments, when atomic diffusion and paraxial diffraction  accurately cancel each other out, depriving the wave phenomenon of a characteristic scale and cancelling diffraction \cite{Firstenberg2009}.

One-dimensional (stripe-shaped) scale-free solitary waves are observed to undergo a form of instability that breaks up \cite{DelRe11}, a process reminiscent of MI. The puzzling fact is that MI is wholly unexpected in a scale-free system, since, as we show below,
the absence of a characteristic spatial scale provides no means for the propagation to
select a dominant perturbation.
Remarkably, in agreement with experiments, we find that the scale-free regime, while curtailing the effects of MI, triggers entirely new beam instabilities.
 
We theoretically investigate the dynamics of Gaussian scale-free one-dimensional solitons in two-dimensional propagation, 
and consider the spatial evolution of a periodic perturbation 
with a transverse size different from that of the pump stripe beam.
Although we find perturbations that can exponentially grow, these depend solely on pump beam waist and not
on pump power, and no preferential spatial period emerges with maximum gain.

We stress that the relevant one-dimensional nonlinear optical model (i.e., when neglecting one trasverse dimension,
as commonly done in MI theory), 
does not admit an unstable regime, and the considered beam breaking is
the result of a subtle multidimensional coupling in the scale-free model, 
such that also the instabilities retain the scale-free character in the fact that no preferential
spatial scale is developed during the dynamics, a feature potentially useful in imaging applications \cite{Barsi2009}.

\noindent {\it Scale-free self-trapped beams ---}
In the paraxial approximation, for a linearly polarized beam with wavelength $\lambda$, the slowly-varying optical field $A$ ($|A|^2=I$ is the optical intensity) obeys the nonlinear propagation equation (hereafter denoted as the {\it the scale-free model}) \cite{DelRe11,Conti11}
\begin{equation}
2  i k \frac{\partial  A}{\partial Z}  +\nabla^2_\perp  A-
\frac{L^2}{\lambda^2} \frac{  (\partial_X I)^2+(\partial_Y I)^2}{4 I^2} A=0\text{,}
\label{soliton_equation_MKS}
\end{equation}
with $k=2\pi n_0/\lambda$, $n_0$ the bulk refractive index, and
the characteristic length $L$ given by 
\begin{equation}
L=4 \pi n_0^2  \epsilon_0 \sqrt{g} \chi_{PNR} (K_B T/q) \text{,}
\end{equation}
where $K_B$ is the Boltzmann constant, $T$ the temperature, 
$q$ the elementary charge, $\epsilon_0$ the vacuum permittivity, $\chi_{PNR}$
the low-frequency permittivity due to the {\it polar nano-regions} (PNR),
and $g>0$ the relevant component of the quadratic electro-opitc tensor.

As has been previously shown \cite{DelRe11, Conti11}, Eq.(\ref{soliton_equation_MKS}), admits bell-shaped soliton solutions 
as long as $L\ge \lambda$. We specifically consider 
a perturbed striped beam.
We begin writing Eq.(\ref{soliton_equation_MKS}) in dimensionless units as
\begin{equation}
  i  \frac{\partial  \psi}{\partial z}  +\frac{1}{2}\nabla^2_{xy}  \psi-
\sigma \frac{  (\partial_x |\psi|^2)^2+(\partial_y|\psi|^2)^2}{|\psi|^4} \psi=0\text{,}
\label{soliton_equation}
\end{equation}
where $x=X/W_0$, $y=Y/W_0$ and $z=Z/Z_0$, with $Z_0=k W_0^2/2$ the diffraction length,
$W_0$ an arbitrary beam waist, and $\sigma=L^2/8\lambda^2$, with $\sigma=1/8$ for $L=\lambda$. In (\ref{soliton_equation}) $\psi=A/A_N$ is the dimensionless field scaled by an arbitrary factor $A_N$ resulting from the intensity-independent feature of the scale-free model.

\noindent The solitary waves of Eq.(\ref{soliton_equation}) 
represent diffraction-free beams with arbitrary intensity and waist \cite{DelRe11,Conti11}. Here we note that Eq.(\ref{soliton_equation}) 
also admits plane wave solutions for $\psi=A_0=\text{constant}$, which, at variance with Kerr media, are not subject to 
any nonlinearly induced phase-shift (i.e., the nonlinear correction to the wave-vector is zero).
We then consider the diffraction-free stripe (i.e., $y-$independent) solutions attained for $\sigma=1/8:$
\begin{equation}
\psi_s(x,z)=\psi_0(x)\exp\left(i\beta z\right)=A_0\exp\left(-\frac{x^2}{w_0^2}\right)\exp\left(i\beta z\right)\text{,}
\label{stripe}
\end{equation}
with $\psi_0(x)=\exp(-x^2/w_0^2)$, $\beta=-1/w_0^2$, and $A_0$ and $w_0$ arbitrary independent parameters, such that this self-trapped beam exists at any intensity level 
and for any waist (within the validity of the paraxial approximation).
Note that as $w_0\rightarrow\infty$, $\beta\rightarrow 0$, one recovers the plane-wave solution $\psi_0=A_0$ mentioned above.
Solutions also exist for $\sigma>1/8$ ($L>\lambda$) \cite{DelRe11,Conti11},
but here we will focus on the case $L\cong\lambda$ ($\sigma=1/8$), as this is 
the condition achieved in experiments.

\noindent {\it Absence of Modulational Instability ---}
We consider perturbations to the exact solution, which is written as
\begin{equation}
\psi(x,y,z)=\left[\psi_0(x)+p(x,y,z)\right]\exp(i\beta z)\text{.}
\label{pert}
\end{equation}
The linearized evolution equation for the perturbation $p$ reads as
\begin{equation}
i\partial_z p +\frac{1}{2}\nabla^2_{xy}p -4\sigma 
\left[ \frac{\psi_{0x}^2}{\psi_0^2}p+\psi_{0x}\left(\frac{p+p^*}{\psi_0}\right)_x\right]=\beta p\text{.}
\label{pert_eq}
\end{equation}
From Eq.(\ref{pert_eq}), one sees that if $\psi_0=A_0$ ($w_0\rightarrow\infty$) no instability is expected 
(i.e., no solution such that $p$ exponentially grows with $z$); the scale-free model of Eq.(\ref{soliton_equation})
is not exhibiting standard MI.
This result is in striking contrast with the well-known fact that MI always accompanies the existence of solitary waves solutions,
as for Kerr, saturable or quadratic nonlinearities. \cite{KivsharBook, TrilloBook}
Conversely, in the scale-free model, one has self-trapped bright beams, but no instability for the plane-wave solutions;
unaffected by MI, the scale-free system turns out to be fertile ground for very different forms of instabilities,
as we discuss below.

\noindent {\it Theory of the scale-free instabilities --- }
We first note that Eq.(\ref{pert_eq}) does not imply an exponential growth 
for a perturbation whose spatial profile in the $x$ direction is the
same as the pump beam $\psi_0$, i.e., for $p$ such that  
\begin{equation}
p(x,y,z)=\psi_0(x) \alpha_+(z) e^{i k_y y}+\psi_0(x) \alpha_-(z)^* e^{-i k_y y}\text{.}
\label{p_psi0}
\end{equation}
Indeed the term containing $p^*$ in Eq.(\ref{pert_eq}) disappears, so that there 
is no instability. Note that as Eq.(\ref{pert_eq}) does not contain coefficients explicitly dependent on $y$, $p$ can be expressed as a plane wave 
expansion with respect to $y$ without loss of generality.
We then write the perturbation as 
\begin{equation}
p(x,y,z)=\psi_1(x) \alpha_+(z) e^{i k_y y}+\psi_1(x) \alpha_-(z)^* e^{-i k_y y}\text{,}
\label{p_psi1}
\end{equation}
with $\psi_1(x)$ an arbitrary spatial profile (specified below), different from $\psi_0(x)$.
To keep the treatment as simple as possible, we limit analysis to the 
Gaussian soliton for $L=\lambda$ ($\sigma=1/8$), and 
we take the profile for the perturbation $\psi_1(x)$ as a Gaussian with waist different from that of the pump $\psi_0(x)$, i.e.,
\begin{equation}
 \psi_1(x)=\left(\frac{2}{\pi w_1^2}\right)^{1/4} \exp(-\frac{x^2}{w_1^2})\text{,}
\end{equation}
such that $\left(\psi_1,\psi_1\right)=1$, with the scalar product $(a,b)=\int a^* b dx$.
We use Eq.(\ref{p_psi1}) in Eq.(\ref{pert_eq}) and project over $\psi_1$, which corresponds to making an expansion in 
a Hermite-Gauss basis  with respect to $x$ and only retaining the first term of the expansion.
We find the coupled equations for the amplitudes $\alpha_\pm$ after Eq.(\ref{p_psi1}):
\begin{equation}
\pm 2 i \frac{d\alpha_\pm}{dz}+
\left(-k_y^2+\frac{1}{w_0^2}-\frac{1}{w_1^2}\right)\alpha_\pm+\frac{w_1^2-w_0^2}{w_0^4}\alpha_\mp=0\text{.}
\label{cmt}
\end{equation}
Note that the last term coupling $\alpha_\pm$ in Eq.(\ref{cmt}) is responsible for the instabilities and is proportional to $w_1^2-w_0^2$, hence for $w_1=w_0$, we recover the result stated above, i.e., the absence of instability for a perturbation with the same $x-$size of the pump beam. Analogously, the instability disappears for $w_0\rightarrow\infty$, corresponding
to the plane-wave case, also discussed above.

\noindent For $w_1\neq w_0$, one finds that (\ref{cmt}) admits exponentially amplified solutions, which are written as 
$\alpha_\pm=\hat{\alpha}_\pm \exp(\lambda z)$, with the gain $\lambda$ given by
\begin{equation}
4\lambda^2(k_y,w_1)=\frac{(w_1^2-w_0^2)^2}{w_0^8}-\left(k_y^2-\frac{1}{w_1^2}+
\frac{1}{w_0^2}\right)^2\text{;}
\end{equation}
with $r=w_1/w_0$, one has
\begin{equation}
4\lambda^2(k_y,r)w_0^4=(r^2-1)^2-\left[(k_y w_0)^2+1-1/r^2\right]^2\text{.}
\label{gain}
\end{equation}
The most unstable perturbation corresponds to the values $r$ and $k_y$ that maximize $\lambda^2$. As detailed below,
the analysis of Eq.(\ref{gain}) identifies two kinds of instabilities: 
with respect to perturbations of width greater than the pump beam ($r>1$), and the opposite case ($r<1$), denoted hereafter 
as {\it defocusing} and {\it fragmenting} instabilities, respectively.

\begin{figure}
\includegraphics[width=\columnwidth]{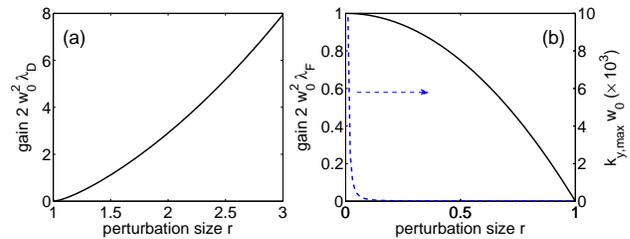}
\caption{ 
(a) Gain versus the ratio of the waist between the 
perturbation and the pump for $r>1$ (maximum gain 
attained at $k_y=0$); (b) Left scale: gain versus $r$ for
$r<1$; right scale: corresponding maximally amplified period.
\label{fig1}}
\end{figure}

\begin{figure}
\includegraphics[width=\columnwidth]{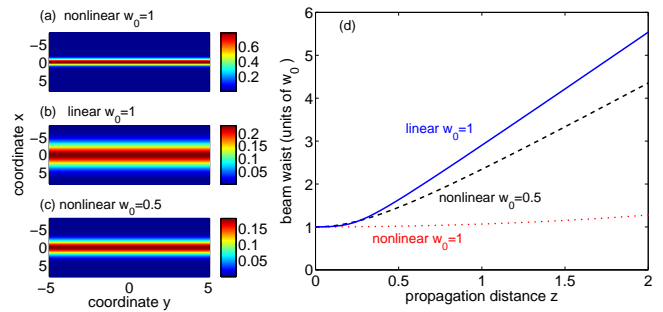}
\caption{
(Color online) 
Simulation of the defocusing instability after Eq.(\ref{soliton_equation}).
(a) Output beam at $z=2$ for $\sigma=0.125$, with a striped beam
with $w_0=1$ and $A_0=1$; (b) as in (a) with $\sigma=0$
(linear propagation);(c) as in (a) with $w_0=0.5$;  (d) beam waist in the $x$ directions for (a), dotted line, for (b), dashed line, for (c) continuous line.
\label{fig2}}
\end{figure}

\noindent {\it Defocusing instability ---}
For a perturbation with $w_1>w_0$, the condition
$\lambda^2>0$ predicts the maximum gain at $k_y=0$ and given by
\begin{equation}
\lambda_D=\frac{\sqrt{(r^2-1)^2-(1-1/r^2)^2}}{2 w_0^2}\text{.}
\label{lambdad}
\end{equation}
$\lambda_D$ is positive only for $r>1$, and is shown in Fig.\ref{fig1}a.
This maximum gain is not limited, and grows with $r$, thus revealing a self-propelling instability, such that if a perturbation with waist greater than the beam is superimposed, the beam tends to spread (the perturbation gains energy) and the spreading rate increases with the waist of the beam.
Note that the gain is maximum at $k_y=0$, denoting an instability that does not tend to alter the striped shape of the beam by introducing periodical modulations. This process is also more pronounced for small waists, as the maximum gain 
$\lambda_D$ goes like $w_0^{-2}$.
We show in figure \ref{fig2}, an example of this instability, as obtained by numerically solving Eq.(\ref{soliton_equation}). The evolution reveals a defocusing of the beam, which is hence unstable, and is compared with the linear case $\sigma=0$ (linear propagation); for $\sigma>0$, the effect is more pronounced 
as the waist $w_0$ is reduced. 
Note that such an instability is not observed for the two dimensional scale-free beams, neither in the numerical simulations nor in the experiments,\cite{DelRe11} and these appear
to be a sort of attractor for the dynamics of the stripe beam, as shown in the following.

\begin{figure}
\includegraphics[width=\columnwidth]{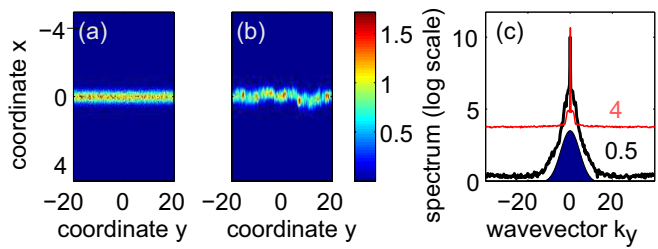}
\caption{
(Color online) Simulation of the fragmenting instability.
The stripe solution is perturbed by a Gaussian noise $p$
with $10\%$ amplitude ($|p/\psi_0|=0.1$); (a) input beam; (b) output beam at $z=2$
($A_0=1$,$w_0=0.5$, $\sigma=0.125$); (c) spectrum obtained after the numerical simulations (continuos thick line for $w_0=0.5$, thin line $w_0=4$), the colored region
is after eq. (\ref{gain}) showing the theoretically expected spectrum.
\label{fig3}}
\end{figure}

\noindent {\it Fragmenting instability ---}
For $r<1$ (perturbation smaller than the pump, i.e., $w_1<w_0$),
the gain is maximum at a $k_y>0$, fixed by $r$, and given by
\begin{equation}
k_{y,max}=\frac{\sqrt{1/r^2-1}}{w_0}\text{.}
\end{equation}
The corresponding maximum growth rate $\lambda_R$ is 
\begin{equation}
\lambda_F=\frac{1-r^2}{2 w_0^2}\text{.}
\end{equation}
However, as $r<1$, the maximum gain corresponds to $r=0$ (vanishing $w_1$) with diverging $k_{y,max}$, denoting the tendency of the beam to break up into very tiny spots, with no preferential spatial scale, in great contrast with the standard MI. Additionally, we note that, for a fixed $r$, the gain scales as the
inverse squared waist, hence the more focused is the
beam, the more pronounced is the instability.
This is another remarkable difference with standard MI;
e.g., in Kerr media as the pump power is increased, the gain grows as well; conversely in the scale-free model, the power does not affect the gain, which, on the contrary, increases when decreasing the beam spot size.
Note also that the gain level for the fragmenting instability
is lower than for the defocusing one and is limited by
the upper value $\lambda_F(r=0)=1/(2w_0^2)$, longer
propagation distances are needed to appreciate its development.
A notable outcome is that tiny details superimposed onto the
pump are amplified upon propagation.
In Fig.\ref{fig3}a,b, we show an example of the fragmenting instability, 
by the evolution of a stripe perturbed by a Gaussian noise with $10\%$ amplitude with respect to the pump.

In Fig.\ref{fig3}c, we show the spectrum obtained numerically from Eq.(\ref{soliton_equation}), averaged over $10$ noise realizations, and compared with that expected from Eq.(\ref{gain});
the shaded region corresponds to the overlap of the amplified spectral regions for various $r$ and multiplied
by a Gaussian spectral content. 
We report spectra for an initial waist $w_0=0.5$ and $w_0=4$, showing that in the latter case the instability is moderated by about an order of magnitude, as expected from Eq.(\ref{gain}). Spectra are arbitrarily shifted in the vertical axis for the sake of comparison.
%%%%%%%%%%%%%%%%%%%%%%%%% figure 1
%%%%%%%%%%%%%%%%%%%%%%%%% figure 2
%%%%%%%%%%%%%%%%%%%%%%%%% figure 3
%%%%%%%%%%%%%%%%%%%%%%%%%%%%%%%%%%%%%%%%%
In recent experiments in copper doped KTN:Li, the described instability has been observed for  with beam waist in the $x-$direction of the order of $15\mu$m, after a propagation distance of $6$mm, corresponding to $W_0=30\mu$m and $w_0=0.5$ in Fig.\ref{fig3}. \cite{DelRe11}

\noindent{\it Conclusions ---} 
We have investigated wave instability in systems supporting 
scale-free optics, such as vitrified photorefractive dipolar glass.
Without a dominant spatial scale, MI is found to disappear, 
outdone by two new forms of instability with a number of 
interesting physical and mathematical properties. 
Scale-free instability is, in fact, able to amplify tiny
beam perturbations at any spatial scale (in the 
paraxial approximation), without washing out or
filtering the noise. The instability hence allows the
wave to pick-up and carry large amounts of information,
a fact that opens interesting perspectives for
imaging applications.

We acknowledge support from the CINECA-ISCRA and Humboldt foundation.
The research leading to these results has received funding from the
European Research Council under the European Community's Seventh Framework Program 
(FP7/2007-2013)/ERC grant agreement n.201766, and from the Italian Ministry of Research (MIUR) through the "Futuro in Ricerca" FIRB-grant PHOCOS - RBFR08E7VA. Partial funding was received
through the SMARTCONFOCAL project of the Regione Lazio. 

%\bibliography{MEGAbib}

\begin{thebibliography}{17}
\expandafter\ifx\csname natexlab\endcsname\relax\def\natexlab#1{#1}\fi
\expandafter\ifx\csname bibnamefont\endcsname\relax
  \def\bibnamefont#1{#1}\fi
\expandafter\ifx\csname bibfnamefont\endcsname\relax
  \def\bibfnamefont#1{#1}\fi
\expandafter\ifx\csname citenamefont\endcsname\relax
  \def\citenamefont#1{#1}\fi
\expandafter\ifx\csname url\endcsname\relax
  \def\url#1{\texttt{#1}}\fi
\expandafter\ifx\csname urlprefix\endcsname\relax\def\urlprefix{URL }\fi
\providecommand{\bibinfo}[2]{#2}
\providecommand{\eprint}[2][]{\url{#2}}

\bibitem[{\citenamefont{Campillo et~al.}(1973)\citenamefont{Campillo, Shapiro,
  and Suydam}}]{Campillo73}
\bibinfo{author}{\bibfnamefont{A.}~\bibnamefont{Campillo}},
  \bibinfo{author}{\bibfnamefont{S.~L.} \bibnamefont{Shapiro}},
  \bibnamefont{and} \bibinfo{author}{\bibfnamefont{B.~R.}
  \bibnamefont{Suydam}}, \bibinfo{journal}{\apl} \textbf{\bibinfo{volume}{23}},
  \bibinfo{pages}{628} (\bibinfo{year}{1973}).

\bibitem[{\citenamefont{Law and G.~A.~Swartzlander}(1993)}]{Law1993}
\bibinfo{author}{\bibfnamefont{C.~T.} \bibnamefont{Law}} \bibnamefont{and}
  \bibinfo{author}{\bibfnamefont{J.}~\bibnamefont{G.~A.~Swartzlander}},
  \bibinfo{journal}{Opt. Lett.} \textbf{\bibinfo{volume}{18}},
  \bibinfo{pages}{586} (\bibinfo{year}{1993}).

\bibitem[{\citenamefont{Mamaev et~al.}(1996)\citenamefont{Mamaev, Saffman, and
  Zozulya}}]{Mamaev96}
\bibinfo{author}{\bibfnamefont{A.~V.} \bibnamefont{Mamaev}},
  \bibinfo{author}{\bibfnamefont{M.}~\bibnamefont{Saffman}}, \bibnamefont{and}
  \bibinfo{author}{\bibfnamefont{A.~A.} \bibnamefont{Zozulya}},
  \bibinfo{journal}{Phys. Rev. Lett.} \textbf{\bibinfo{volume}{76}},
  \bibinfo{pages}{2262} (\bibinfo{year}{1996}).

\bibitem[{\citenamefont{Fuerst et~al.}(1997)\citenamefont{Fuerst, Baboiu,
  Lawrence, Torruellas, Stegeman, Trillo, and Wabnitz}}]{Fuerst97}
\bibinfo{author}{\bibfnamefont{R.~A.} \bibnamefont{Fuerst}},
  \bibinfo{author}{\bibfnamefont{D.-M.} \bibnamefont{Baboiu}},
  \bibinfo{author}{\bibfnamefont{B.}~\bibnamefont{Lawrence}},
  \bibinfo{author}{\bibfnamefont{W.~E.} \bibnamefont{Torruellas}},
  \bibinfo{author}{\bibfnamefont{G.~I.} \bibnamefont{Stegeman}},
  \bibinfo{author}{\bibfnamefont{S.}~\bibnamefont{Trillo}}, \bibnamefont{and}
  \bibinfo{author}{\bibfnamefont{S.}~\bibnamefont{Wabnitz}},
  \bibinfo{journal}{Phys. Rev. Lett.} \textbf{\bibinfo{volume}{78}},
  \bibinfo{pages}{2756} (\bibinfo{year}{1997}).

\bibitem[{\citenamefont{Schiek et~al.}(2001)\citenamefont{Schiek, Fang,
  Malendevich, and Stegeman}}]{Shiek01}
\bibinfo{author}{\bibfnamefont{R.}~\bibnamefont{Schiek}},
  \bibinfo{author}{\bibfnamefont{H.}~\bibnamefont{Fang}},
  \bibinfo{author}{\bibfnamefont{R.}~\bibnamefont{Malendevich}},
  \bibnamefont{and} \bibinfo{author}{\bibfnamefont{G.~I.}
  \bibnamefont{Stegeman}}, \bibinfo{journal}{Phys. Rev. Lett.}
  \textbf{\bibinfo{volume}{86}}, \bibinfo{pages}{4528} (\bibinfo{year}{2001}).

\bibitem[{\citenamefont{Peccianti et~al.}(2003)\citenamefont{Peccianti, Conti,
  and Assanto}}]{Peccianti03}
\bibinfo{author}{\bibfnamefont{M.}~\bibnamefont{Peccianti}},
  \bibinfo{author}{\bibfnamefont{C.}~\bibnamefont{Conti}}, \bibnamefont{and}
  \bibinfo{author}{\bibfnamefont{G.}~\bibnamefont{Assanto}},
  \bibinfo{journal}{\pre} \textbf{\bibinfo{volume}{68}},
  \bibinfo{pages}{025602(R)} (\bibinfo{year}{2003}).

\bibitem[{\citenamefont{DelRe and Segev}(2009)}]{DelRe09book}
\bibinfo{author}{\bibfnamefont{E.}~\bibnamefont{DelRe}} \bibnamefont{and}
  \bibinfo{author}{\bibfnamefont{M.}~\bibnamefont{Segev}},
  \emph{\bibinfo{title}{Self-Focusing and Solitons in Photorefractive Media}}
  (\bibinfo{publisher}{Springer}, \bibinfo{address}{Berlin},
  \bibinfo{year}{2009}), vol. \bibinfo{volume}{114} of
  \emph{\bibinfo{series}{Topics in Applied Physics}}, pp.
  \bibinfo{pages}{547--572}.

\bibitem[{\citenamefont{Kivshar and Agrawal}(2003)}]{KivsharBook}
\bibinfo{author}{\bibfnamefont{Y.~S.} \bibnamefont{Kivshar}} \bibnamefont{and}
  \bibinfo{author}{\bibfnamefont{G.~P.} \bibnamefont{Agrawal}},
  \emph{\bibinfo{title}{Optical solitons}} (\bibinfo{publisher}{Academic
  Press}, \bibinfo{address}{New York}, \bibinfo{year}{2003}).

\bibitem[{\citenamefont{Trillo and Torruealls}(2001)}]{TrilloBook}
\bibinfo{editor}{\bibfnamefont{S.}~\bibnamefont{Trillo}} \bibnamefont{and}
  \bibinfo{editor}{\bibfnamefont{W.}~\bibnamefont{Torruealls}}, eds.,
  \emph{\bibinfo{title}{Spatial Solitons}}
  (\bibinfo{publisher}{Springer-Verlag}, \bibinfo{address}{Berlin},
  \bibinfo{year}{2001}).

\bibitem[{\citenamefont{DelRe et~al.}(2011)\citenamefont{DelRe, Spinozzi,
  Agranat, and Conti}}]{DelRe11}
\bibinfo{author}{\bibfnamefont{E.}~\bibnamefont{DelRe}},
  \bibinfo{author}{\bibfnamefont{E.}~\bibnamefont{Spinozzi}},
  \bibinfo{author}{\bibfnamefont{R.}~\bibnamefont{Agranat}}, \bibnamefont{and}
  \bibinfo{author}{\bibfnamefont{C.}~\bibnamefont{Conti}},
  \bibinfo{journal}{Nature Photonics} \textbf{\bibinfo{volume}{5}},
  \bibinfo{pages}{39} (\bibinfo{year}{2011}).

\bibitem[{\citenamefont{{Conti} et~al.}(2011)\citenamefont{{Conti}, {Agranat},
  and {DelRe}}}]{Conti11}
\bibinfo{author}{\bibfnamefont{C.}~\bibnamefont{{Conti}}},
  \bibinfo{author}{\bibfnamefont{A.~J.} \bibnamefont{{Agranat}}},
  \bibnamefont{and} \bibinfo{author}{\bibfnamefont{E.}~\bibnamefont{{DelRe}}},
  \bibinfo{journal}{ArXiv e-prints}  (\bibinfo{year}{2011}),
  \eprint{1102.4945}.

\bibitem[{\citenamefont{Crosignani et~al.}(1999)\citenamefont{Crosignani,
  Degasperis, DelRe, {Di Porto}, and Agranat}}]{Crosignani99}
\bibinfo{author}{\bibfnamefont{B.}~\bibnamefont{Crosignani}},
  \bibinfo{author}{\bibfnamefont{A.}~\bibnamefont{Degasperis}},
  \bibinfo{author}{\bibfnamefont{E.}~\bibnamefont{DelRe}},
  \bibinfo{author}{\bibfnamefont{P.}~\bibnamefont{{Di Porto}}},
  \bibnamefont{and} \bibinfo{author}{\bibfnamefont{A.~J.}
  \bibnamefont{Agranat}}, \bibinfo{journal}{Phys. Rev. Lett.}
  \textbf{\bibinfo{volume}{82}}, \bibinfo{pages}{1664} (\bibinfo{year}{1999}).

\bibitem[{\citenamefont{Chauvet~M and G}(2006)}]{Chauvet2006}
\bibinfo{author}{\bibfnamefont{F.~G.~Y.} \bibnamefont{Chauvet~M},
  \bibfnamefont{Guo A~Q}} \bibnamefont{and}
  \bibinfo{author}{\bibfnamefont{S.}~\bibnamefont{G}}, \bibinfo{journal}{J.
  Appl. Phys.} \textbf{\bibinfo{volume}{99}}, \bibinfo{pages}{113107}
  (\bibinfo{year}{2006}).

\bibitem[{\citenamefont{Asaro et~al.}(2005)\citenamefont{Asaro, Sheldon, Chen,
  Ostroverkhova, and Moerner}}]{Asaro2005}
\bibinfo{author}{\bibfnamefont{M.}~\bibnamefont{Asaro}},
  \bibinfo{author}{\bibfnamefont{M.}~\bibnamefont{Sheldon}},
  \bibinfo{author}{\bibfnamefont{Z.~G.} \bibnamefont{Chen}},
  \bibinfo{author}{\bibfnamefont{O.}~\bibnamefont{Ostroverkhova}},
  \bibnamefont{and} \bibinfo{author}{\bibfnamefont{W.~E.}
  \bibnamefont{Moerner}}, \bibinfo{journal}{Opt. Lett.}
  \textbf{\bibinfo{volume}{30}}, \bibinfo{pages}{519} (\bibinfo{year}{2005}).

\bibitem[{\citenamefont{Sheu and Shih}(2001)}]{Sheu2001}
\bibinfo{author}{\bibfnamefont{F.}~\bibnamefont{Sheu}} \bibnamefont{and}
  \bibinfo{author}{\bibfnamefont{M.}~\bibnamefont{Shih}}, \bibinfo{journal}{J.
  Opt. Soc. Am. B} \textbf{\bibinfo{volume}{18}}, \bibinfo{pages}{785}
  (\bibinfo{year}{2001}).

\bibitem[{\citenamefont{Firstenberg et~al.}(2009)\citenamefont{Firstenberg,
  London, Shuker, Ron, and Davidson}}]{Firstenberg2009}
\bibinfo{author}{\bibfnamefont{O.}~\bibnamefont{Firstenberg}},
  \bibinfo{author}{\bibfnamefont{P.}~\bibnamefont{London}},
  \bibinfo{author}{\bibfnamefont{M.}~\bibnamefont{Shuker}},
  \bibinfo{author}{\bibfnamefont{A.}~\bibnamefont{Ron}}, \bibnamefont{and}
  \bibinfo{author}{\bibfnamefont{N.}~\bibnamefont{Davidson}},
  \bibinfo{journal}{Nat. Phys.} \textbf{\bibinfo{volume}{5}},
  \bibinfo{pages}{665} (\bibinfo{year}{2009}).

\bibitem[{\citenamefont{Barsi et~al.}(2009)\citenamefont{Barsi, Wan, and
  Fleischer}}]{Barsi2009}
\bibinfo{author}{\bibfnamefont{C.}~\bibnamefont{Barsi}},
  \bibinfo{author}{\bibfnamefont{W.}~\bibnamefont{Wan}}, \bibnamefont{and}
  \bibinfo{author}{\bibfnamefont{J.~W.} \bibnamefont{Fleischer}},
  \bibinfo{journal}{Nature Photonics} \textbf{\bibinfo{volume}{3}},
  \bibinfo{pages}{211} (\bibinfo{year}{2009}).

\end{thebibliography}

\end{document}